%% file: PRL_Bhh_BR_V30.tex
 \documentclass[aps,prl,twocolumn,amsmath,amssymb,superscriptaddress,showpacs]{revtex4}

\usepackage{graphicx}
\usepackage{dcolumn}
\usepackage{bm}
\usepackage[abs]{overpic} 
\preprint{draftV3.1}

   \newcommand{\ppbar}{\ensuremath{\bar{p}p}}                                                                                             %
   \newcommand{\Bd}{\ensuremath{B^{0}}}                                                                                                   %
   \newcommand{\Bu}{\ensuremath{B^{+}}}                                                                                                   %
   \newcommand{\Bs}{\ensuremath{B_{s}^{0}}}                                                                                               %
   \newcommand{\Bdpipi}{\ensuremath{\Bd\rightarrow\pi^{+}\pi^{-}}}                                                                       %
   \newcommand{\BdKpi}{\ensuremath{\Bd\rightarrow K^{+}\pi^{-}}}                                                                        %
   \newcommand{\BdKK}{\ensuremath{\Bd\rightarrow K^{+}K^{-}}}                                                                           %
   \newcommand{\BsKK}{\ensuremath{\Bs\rightarrow K^{+}K^{-}}}                                                                           %
   \newcommand{\BsKpi}{\ensuremath{\Bs\rightarrow K^{-}\pi^{+}}}                                                                        %
   \newcommand{\Bspipi}{\ensuremath{\Bs\rightarrow\pi^{+}\pi^{-}}}                       
   \newcommand{\Lbppi}{\ensuremath{\Lb \rightarrow p\pi^{-}}}   
   \newcommand{\LbpK}{\ensuremath{\Lb \rightarrow p K^{-}}}

   \newcommand{\DKpi}{\ensuremath{D^0\rightarrow K^-\pi^+}}             
   \newcommand{\Lb}{\ensuremath{\Lambda^0_b}}           
   \newcommand{\BsKpisuBdKpidef}{\ensuremath{\frac{f_s}{f_d}\frac{\BR(\Bs\rightarrow K^-\pi^+)}{\BR(\Bd\rightarrow K^+\pi^-)}}}
   \newcommand{\BspipisuBdKpidef}{\ensuremath{\frac{f_s}{f_d}\frac{\BR(\Bs\rightarrow \pi^-\pi^+)}{\BR(\Bd\rightarrow K^+\pi^-)}}}
   \newcommand{\BdKKsuBdKpidef}{\ensuremath{\frac{\BR(\Bd\rightarrow K^-K^+)}{\BR(\Bd\rightarrow K^+\pi^-)}}}
   \newcommand{\LbppisuBdKpidef}{\ensuremath{\frac{\mathit{f_{\Lambda}}}{\mathit{f_d}}\frac{\BR(\Lbppi)}{\BR(\BdKpi)}}}
   \newcommand{\LbpKsuBdKpidef}{\ensuremath{\frac{\mathit{f_{\Lambda}}}{\mathit{f_d}}\frac{\BR(\LbpK)}{\BR(\BdKpi)}}}  
   \newcommand{\BR}{\ensuremath{\mathcal B}}                                                                                   %
   \newcommand{\dedx}{\ensuremath{\mathit{dE/dx}}}                                                                              %
   \newcommand{\micron}{\ensuremath{\mu\mathrm{m}}}
   \newcommand{\pgev}{\ensuremath{\mathrm{GeV}/c}}
   \newcommand{\ptot}{\ensuremath{p_{\mathit{tot}}}}   
   \newcommand{\massgev}{\ensuremath{\mathrm{GeV}/c^{2}}}
   \newcommand{\massmev}{\ensuremath{\mathrm{MeV}/c^{2}}}

\newcommand{\lumifb}{ fb$^{-1}$}
\newcommand{\like}{\ensuremath{\mathcal{L}}}

\newcommand{\mpipi}{\ensuremath{m_{\pipi}}}
\newcommand{\pipi}{\ensuremath{\pi\pi}}
\newcommand{\stat}{\ensuremath{\mathrm{~(stat)}}}		
\newcommand{\syst}{\ensuremath{\mathrm{~(syst)}}}		

\newcommand{\Lxy}{\ensuremath{L_{T}}}			
\begin{document}
\title{Observation of New Charmless Decays of Bottom Hadrons}
\input{July_2008_Authors_Dec22_Rick}
\date{September 21, 2008}
\begin{abstract}
We search for new charmless decays of neutral $b$--hadrons to pairs of 
charged hadrons with the upgraded Collider Detector at the Fermilab Tevatron.
Using a data sample corresponding to \mbox{1\lumifb} of integrated luminosity, 
we report the first observation of the \BsKpi\ decay, with a significance of 
$8.2\sigma$, and measure 
$\BR(\BsKpi)= (5.0 \pm 0.7\stat \pm 0.8\syst)\times 10^{-6}$. 
We also report the first observation of charmless $b$--baryon decays
in the channels \Lbppi\ and \LbpK\ with significances
of $6.0\sigma$ and $11.5\sigma$ respectively, and we measure 
$\BR(\Lbppi) = (3.5 \pm 0.6\stat \pm 0.9\syst)\times 10^{-6}$ and 
$\BR(\LbpK) = (5.6 \pm 0.8\stat \pm 1.5\syst)\times 10^{-6}$. 
No evidence is found for the decays \BdKK\ and \Bspipi, and we set an 
improved upper limit $\BR(\Bspipi) < 1.2\times 10^{-6}$ at the 90\% confidence level.
All quoted branching fractions are measured using $\BR(\BdKpi)$ as a 
reference.
\end{abstract}

\maketitle


Two-body non-leptonic charmless decays of $b$--hadrons are among the most 
widely studied processes in flavor physics. 
The variety of open channels involving similar final states provides 
crucial experimental information to improve the accuracy of 
effective models of strong interaction dynamics. 
The quark-level transition
$b\to u$ makes decay amplitudes sensitive to $\gamma$, the least known
angle of the quark-mixing (Cabibbo-Kobayashi-Maskawa, CKM) matrix.
Significant contributions from higher-order (`penguin') transitions
provide sensitivity to the possible presence of new physics in internal
loops, if the observed decay rates are inconsistent with expectations.

Rich experimental data are currently available for \Bu\ and \Bd\ mesons, 
produced in large quantities in $\Upsilon(4S)$ 
decays~\cite{Aubert:2006fha},  
while much less is experimentally known about the charmless decay modes of the 
\Bs , which are expected to exhibit an equally rich phenomenology.
Information from \Bs\ decays is needed to
better constrain the phenomenological models
of hadronic amplitudes in heavy flavor decays. 
This would lead to increased precision in comparing data to predictions, 
allowing extraction of CKM parameters from non-tree-level 
amplitudes~\cite{Fleischer:1999pa} 
and greater sensitivity to new physics contributions.

Of the possible \Bs\ decay modes into pairs of charmless 
pseudoscalar mesons, only the \BsKK\ has been
observed to date~\cite{Abulencia:2006psa}. The \BsKpi\
is of particular interest, 
because its branching fraction is
sensitive to the CKM angle $\gamma$~\cite{Gronau:2000md} and 
the current experimental bound~\cite{Abulencia:2006psa} is lower than most
predictions~\cite{Sun:2002rn,Beneke:2003zv,Ali:2007ff}.                                                        

A measurement of the branching fraction of the \Bspipi\ mode, along 
with the \BdKK\ mode, would allow a determination of the strength of 
penguin-annihilation amplitudes~\cite{Buras:2004ub}, which is
 currently poorly known and a source of significant uncertainty in
 many calculations~\cite{Beneke:2003zv}. The present search is
 sensitive to both modes.
Two--body charmless decays are also expected from bottom baryons. 
The modes \LbpK\ and \Lbppi\ are predicted to have measurable branching 
fractions, of order $10^{-6}$~\cite{Mohanta:2000nk}, and, in
addition to the interest in their observation, must be considered
as a possible background to the rare \Bs\ and \Bd\ modes being investigated.


In this Letter we report the results of a
search for rare decays of neutral bottom hadrons into a pair of charged
charmless hadrons ($p$, $K$ or $\pi$), performed in 1\lumifb\ of \ppbar\ 
collisions at $\sqrt{s} = 1.96$ TeV, collected 
by the upgraded Collider Detector (CDF II) at the Fermilab Tevatron.
We report the first observation of modes \BsKpi, \LbpK, and 
\Lbppi, and measure their relative branching fractions~\cite{C-conjugate}.

CDF II is a multipurpose magnetic spectrometer surrounded by
calorimeters and muon detectors. 
The detector components relevant for this analysis are briefly 
outlined below; a more detailed 
description can be found in Ref.~\cite{Acosta:2004yw}.
A silicon microstrip vertex detector (SVX) 
and a cylindrical drift chamber (COT) 
immersed in a 1.4~T axial magnetic field
allow reconstruction of charged--particle trajectories (tracks) in the pseudorapidity range 
$|\eta|< 1.0$~\cite{CDF-coordinates}. 
The SVX consists of six concentric layers of double-sided silicon sensors 
with radii between 2.5 and 22 cm, each providing a measurement with up to 15 (70) \micron\ resolution in the 
$\phi$ ($z$) direction.
The COT has 96 measurement layers, between 40 and 137 cm in radius, organized into 
alternating axial and $\pm 2^{\circ}$ stereo superlayers.
The transverse momentum resolution is $\sigma_{p_{T}}/p_{T}^2 \sim 
0.15\%/(\pgev)$, corresponding to a typical mass resolution of 22~\massmev\ for our signals.   
The specific ionization energy loss (\dedx) of charged particles in the COT can
be measured from the collected charge, which is logarithmically encoded in the output 
pulse width of each wire, 
and provides $1.5\sigma$ separation between kaons and pions with momenta 
greater than $2~\pgev$. 

The data were collected by a three-level 
trigger system, using a set of requirements specifically aimed at selecting
two-pronged $B$ decays.
At level~1, COT tracks are reconstructed in the transverse plane
by a hardware processor (XFT)~\cite{Thomson:2002xp}. 
Two opposite-charge particles are required, 
with reconstructed transverse momenta 
$p_{T1}, p_{T2} > 2~\pgev$, the scalar sum $p_{T1}+p_{T2} > 5.5~\pgev$, and an 
azimuthal opening-angle $\Delta\phi < 135^{\circ}$.
At level~2, the silicon vertex trigger (SVT)~\cite{Ashmanskas:2003gf} 
combines XFT tracks with SVX hits
to measure the impact parameter $d$ 
(distance of closest approach to the beam line)
of each track with 45 \micron\ resolution. The requirement of two tracks with $0.1 < 
d < 1.0$ mm reduces the light quark background by two orders of magnitude while 
preserving about half of the signal. 
A tighter opening-angle requirement, $20^{\circ} < \Delta\phi < 
135^{\circ}$, preferentially selects two--body $B$ decays over multi--body decays with $97\%$ efficiency 
and further reduces background.
Each track pair is then used to form a $B$ candidate, which is 
required to have an impact parameter $d_B < 140~\micron$ and to have travelled a distance $\Lxy > 200~\micron$
in the transverse plane.
At level~3, an array of computers confirms the selection with a full event reconstruction.
The overall acceptance of the trigger selection is $\approx 2\% $ for 
$b$--hadrons with $p_{T}>4~\pgev$ and $|\eta| < 1$.  

The offline selection is based on a more accurate 
determination of the same quantities used in the trigger, with the 
addition of two further observables: 
the isolation ($I_{B}$) of the $B$ candidate~\cite{Isolation},
and the quality of the three-dimensional fit ($\chi^{2}$ with 1 d.o.f.) of the 
decay vertex of the $B$ candidate. 
Requiring a large value of $I_{B}$ reduces the background from 
light-quark jets, and a low $\chi^{2}$ reduces the 
background from decays of different long-lived particles within the event, 
owing to the good resolution of the SVX detector in the $z$ direction.
The selection is optimized for detection of the
\BsKpi\ mode.
Maximal sensitivity for both discovery and limit setting
is achieved with a single choice of selection requirements~\cite{Punzi:2003bu} by  minimizing the variance of the estimate of the branching  fraction
in the absence of signal 
\cite{Morello-Thesis}.
The variance is evaluated by performing the full measurement
procedure on simulated samples containing background and all signals  from
the known modes, but no \BsKpi\ signal.
The background fraction for each selection is determined from data
by extrapolating the mass sidebands of the signal, and the signal
yield is predicted by a detailed detector simulation.
This procedure yields the final selection: $I_B>0.525$, $\chi^{2}<5$,
$d > 120~\micron$, $d_B < 60~\micron$, and $\Lxy > 350~\micron$.

\begin{figure}[htb]
\includegraphics[scale=0.35]{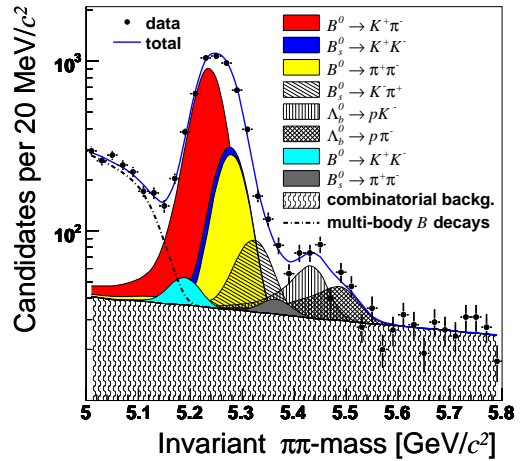}
\caption{\label{fig:projections} Mass distribution
of reconstructed candidates. The charged pion mass is assigned to both tracks.
The total projection and projections of each signal and background 
component of the likelihood fit are overlaid on 
the data distribution. Signals and multi-body $B$ background components are shown stacked on the combinatorial 
background component.}
\end{figure}

No more than one $B$ candidate per event is found after this 
selection, and a mass ($m_{\pi\pi}$) is assigned to 
each, using a charged pion mass assignment for both decay products. 
The resulting mass distribution is shown in Fig.~\ref{fig:projections}. A large peak is visible, 
dominated by the overlapping contributions of the \BdKpi, \Bdpipi, and \BsKK\ 
modes~\cite{Abulencia:2006psa}. 
A \BdKK\ signal would appear as an enhancement around $5.18~\massgev$,
while signals for the other modes of this search are expected at
masses higher than the main peak (5.33--5.55 \massgev). Backgrounds 
include mis-reconstructed multi-body $b$--hadron decays (physics background) and random pairs of charged 
particles (combinatorial background).

We used an unbinned likelihood fit, incorporating kinematic (kin) 
and particle identification (PID) information, to determine the 
fraction of each individual mode in our sample.
The likelihood for the $i$th event is
\begin{eqnarray}\label{eq:likelihood}
    \mathcal{L}_i & = & (1-b)\sum_{j} f_j \mathcal{L}^{\mathrm{kin}}_j  \mathcal{L}^{\mathrm{PID}}_j \nonumber \\
                  &   & +  b \left( f_{\rm{p}} \mathcal{L}^{\mathrm{kin}}_{\mathrm{p}}
    \mathcal{L}^{\mathrm{PID}}_{\mathrm{p}}+
   (1-f_{\rm{p}}) \mathcal{L}^{\mathrm{kin}}_{\mathrm{c}}
    \mathcal{L}^{\mathrm{PID}}_{\mathrm{c}}
	\right),
\end{eqnarray}
where the index $j$ runs over all signal modes, 
 and the index `p' (`c') labels the physics (combinatorial) 
background terms. The $f_j$ are the signal fractions to be determined by 
the fit, together with the background fraction parameters $b$ and $f_{\rm{p}}$.

\begin{figure*}[htb]
\begin{overpic}[scale=0.29]{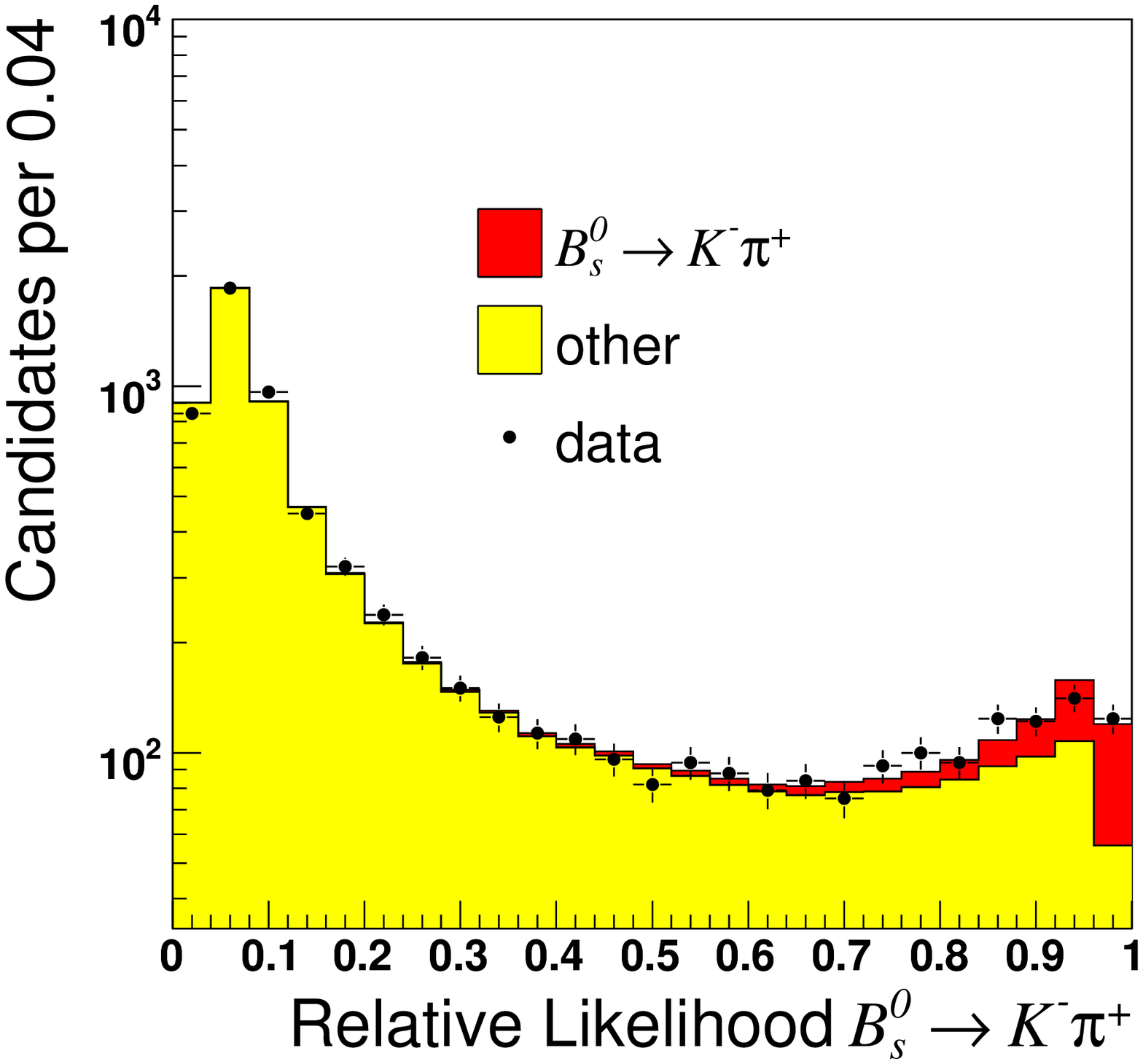}
\put(140,130){(a)}
\end{overpic}   
\hspace{0.05cm}
\begin{overpic}[scale=0.29]{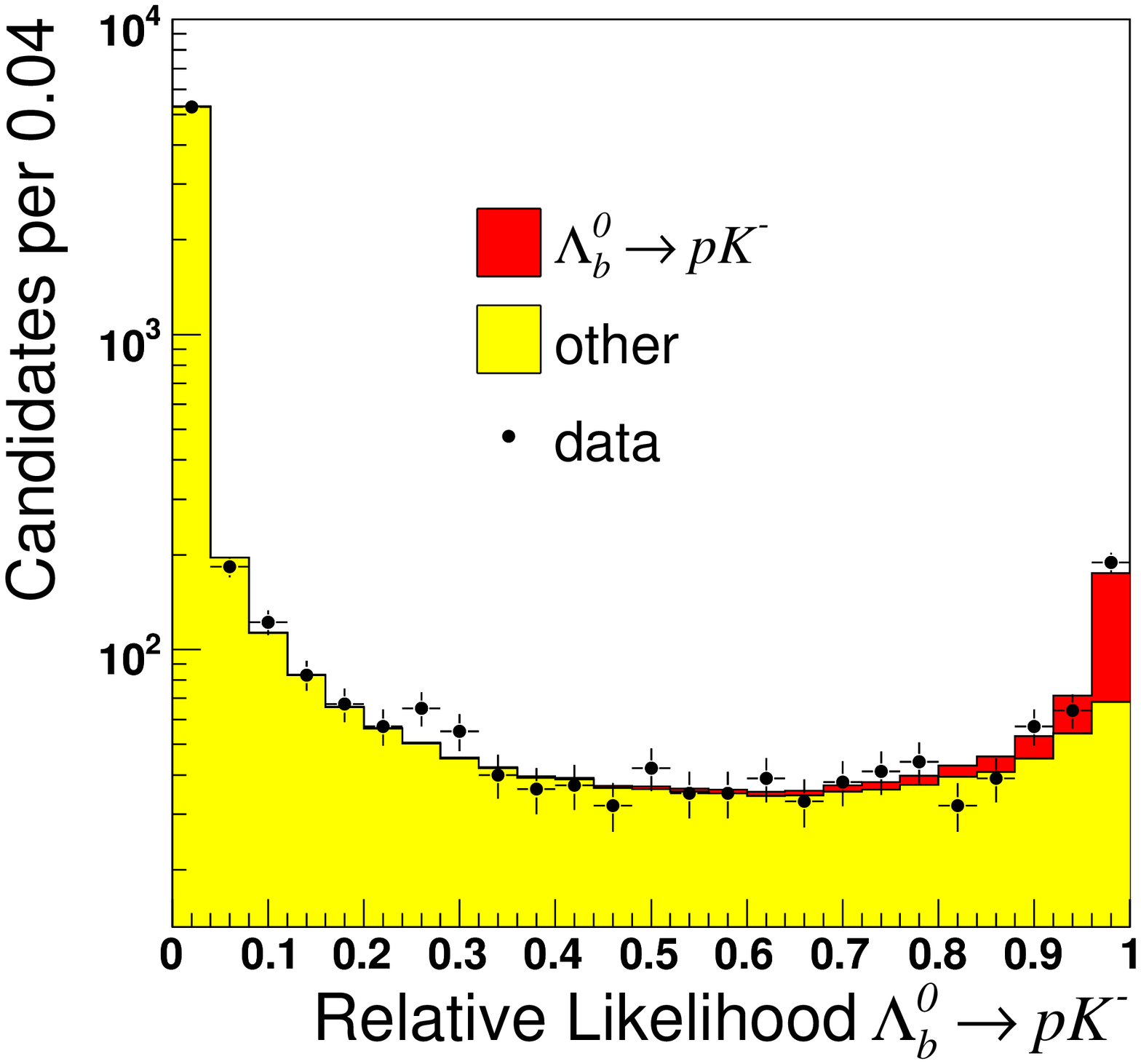}
\put(140,130){(b)}
\end{overpic}   
\hspace{0.05cm}
\begin{overpic}[scale=0.29]{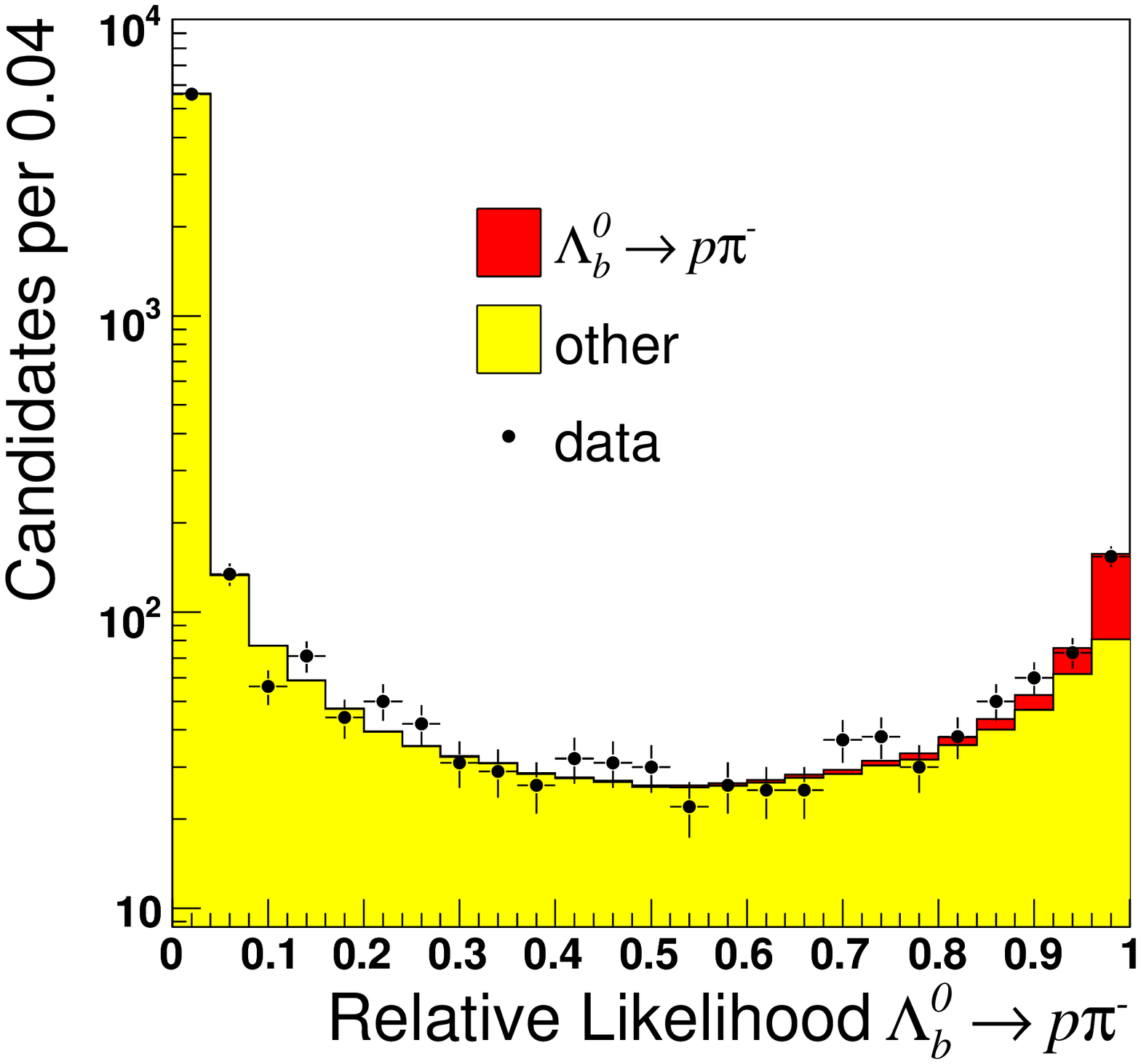}
\put(140,130){(c)}
\end{overpic}  
\caption{Distribution of the relative signal likelihood, $\like_{S}/(\like_{S}+\like_{{\rm other}})$, in the region 
$5.1 <\mpipi< 5.6~\massgev$. For each event, $\like_{S}$ is the 
likelihood for the \BsKpi\ (a), \LbpK\ (b), or \Lbppi\ (c) signal 
hypotheses, and $\like_{\rm other}$ is the likelihood for everything 
but the chosen signal, i.e. the weighted combination of all other 
components according to their measured fractions. Points with error 
bars show the distributions of data and histograms show the distributions predicted from the measured fractions.}
\label{fig:LRplots}
\end{figure*}

The kinematic information is summarized by three loosely correlated
observables: (a) the mass
\mpipi ; (b) the signed momentum imbalance
$\alpha = (1-p_1/p_2) q_{1}$, where $p_1$ ($p_2$) is the
lower (higher) of the particle momenta, and $q_1$ is the sign of the charge of the
particle of momentum $p_{1}$; (c) the scalar sum of particle momenta 
$\ptot=p_1 + p_2$.
The above variables allow evaluation of the invariant mass $m_{12}$ of a candidate for 
any mass assignment of the decay products ($m_{1}$,$m_{2}$), using the equation
\begin{eqnarray}\label{eq:Mpipi2}
 m^{2}_{12}  =  m^{2}_{\pi\pi}  -  2 m_{\pi}^2 + m_{1}^2+m_{2}^2 +
                 \nonumber    \\
          -  2 \sqrt{p_{1}^2+m_{\pi}^2} \sqrt{p_{2}^2+m_{\pi}^2}  
                    +  2\sqrt{p_{1}^2+m_{1}^2} \sqrt{p_{2}^2+m_{2}^2},
 \end{eqnarray}

 where
$p_1 = \frac{1-|\alpha|}{2-|\alpha|}\ptot$ , $p_2 = 
 \frac{1}{2-|\alpha|}\ptot$. 

We used the mass sidebands in data $(\mpipi \in [5.00,5.12]\cup [5.6,6.2]~\massgev)$ 
to obtain the kinematic distributions of backgrounds~\cite{Morello-Thesis}.
The mass distribution of the combinatorial background is 
parameterized by an exponential function, while the physics background is modeled by an 
ARGUS function~\cite{Albrecht:1990am} 
convoluted with a Gaussian resolution function. 
In order to ensure the reliability of the search for small signals in 
the vicinity of larger peaks, the shapes of the mass distributions assigned to each signal have been 
modeled in detail. We have included
the momentum dependence and non--Gaussian tails of resolution from a full 
simulation of the detector, and the effects of soft photon 
radiation in the final state, based on recent QED 
calculations~\cite{Baracchini:2005wp}.
This resolution model was checked against the observed shape 
of the \DKpi\ signal in a sample of $1.5\times 10^{6}$ $D^{*+}\to 
D^0\pi^+$ decays, collected with a similar trigger selection. 
The observed discrepancies are below the 
$10^{-3}$ level, and their effect on the present measurement is negligible in 
comparison with other systematic uncertainties.
The $D^{*+}\to D^0\pi^+$ sample was also used to calibrate the \dedx\ response of the drift chamber 
to kaons and pions, using the charge of the $D^{*+}$ pion to identify the $D^0$ decay
products. The \dedx\ response of protons was determined from a sample of about 124,000 
$\Lambda^{0}\to p \pi^{-}$ decays.
The model of the background allows for pion, kaon, proton, and electron components, 
whose fractions are determined by the fit. Muons are indistinguishable from pions with the
available 10\% fractional \dedx\ resolution and are therefore incorporated into the 
pion component.


From the signal fractions returned by the likelihood fit we 
calculate the signal yields shown in Table~\ref{tab:yields}.
The significance of each signal is evaluated as the ratio of the 
yield observed in data, and its total uncertainty (statistical and systematic)
as determined from a simulation where the size of that signal is set to zero. 
This evaluation assumes a Gaussian distribution of yield estimates, 
supported by the results obtained from repeated fits to simulated samples.
This procedure yields a more accurate measure of significance with 
respect to the purely statistical estimate obtained from $\sqrt{-2\Delta {\rm ln}({\cal 
L})}$.
We obtain significant signals for the \BsKpi\ mode ($8.2\sigma$), 
and for the \Lbppi\ ($6.0\sigma$) and \LbpK\ ($11.5\sigma$) modes.
Figure~\ref{fig:LRplots} shows relative likelihood distributions for 
these modes.
No evidence is found for the modes \Bspipi\ or \BdKK , in agreement 
with expectations of significantly smaller branching fractions. 

\begin{table}
\caption{\label{tab:yields} Yields and significances of rare mode 
signals.
The first quoted uncertainty is statistical, the second is 
systematic.} 

{
\begin{tabular}{lcc}
\hline \hline
Mode 	& N$_{s}$	&  Significance 			\\
\hline
\BsKpi	& 230 $\pm$ 34 $\pm$ 16	& $8.2 \sigma$  		\\
\Bspipi	& 26 $\pm$  16 $\pm$ 14	& $<3\sigma$	\\
\BdKK	& 61 $\pm$  25 $\pm$ 35	& $<3\sigma$	\\
\LbpK	& 156 $\pm$ 20 $\pm$ 11	& $11.5\sigma$		\\
\Lbppi	& 110 $\pm$ 18 $\pm$ 16	& $6.0 \sigma$          \\
\hline \hline
\end{tabular}
}
\end{table}
\begin{table*}[htb]
\caption{\label{tab:BRs} Measured relative branching fractions of
rare modes. The ratio $f_{\Lambda}/f_{d}$ is $p_T$--dependent~\cite{Aaltonen:2008eu}, and is 
defined here as: $f_{\Lambda}/f_{d} = \sigma(p\bar{p}\to\Lb X;p_{T} > 6~\pgev,|\eta|<1)/\sigma(p\bar{p}\to\Bd X;
p_{T} > 6~\pgev,|\eta|<1)$.
Absolute branching fractions were derived by normalizing to the current world--average value
${\mathcal B}(\mbox{\BdKpi}) = (19.4\pm 0.6) \times 10^{-6}$, and assuming the average values at high energy for the production fractions: 
$f_{s}/f_{d}= 0.276\pm 0.034$, and $f_{\Lambda}/f_{d} = 0.230 \pm 0.052$ \cite{PDG08}.
The first quoted uncertainty is statistical, the second is
systematic.}

{
\begin{tabular}{lcrclcc}
\hline\hline
Mode && \multicolumn{3}{c}{Relative \BR}   && Absolute \BR ($10^{-6}$)\\
\hline
\BsKpi	&& \BsKpisuBdKpidef    &=&       0.071 $\pm$ 0.010 
$\pm$ 0.007        && 5.0 $\pm$ 0.7 $\pm$ 0.8   \\
\Bspipi	&& \BspipisuBdKpidef   &=&       0.007 $\pm$ 0.004 
$\pm$ 0.005        && 0.49 $\pm$ 0.28 $\pm$ 0.36 ($<1.2$ at 90\% C.L.)\\
\BdKK	&& \BdKKsuBdKpidef   &=&       0.020 $\pm$ 0.008 $\pm$ 
0.006          && 0.39 $\pm$ 0.16 $\pm$ 0.12 ($<0.7$ at 90\% C.L.)\\
\LbpK	&& \LbpKsuBdKpidef    &=&       0.066 $\pm$ 0.009 $\pm$ 
0.008        && 5.6 $\pm$ 0.8 $\pm$ 1.5   \\
\Lbppi	&&  \LbppisuBdKpidef   &=&        0.042 $\pm$ 0.007 
$\pm$ 0.006      && 3.5 $\pm$ 0.6 $\pm$ 0.9      \\
\hline\hline
\end{tabular}
}
\end{table*}


To avoid large uncertainties associated with production cross sections 
and absolute reconstruction efficiency, we measure all branching 
fractions relative to the \BdKpi\ mode.
Frequentist upper limits~\cite{Feldman:1997qc} at the 90\% C.L. are quoted for the unseen modes.
For the measurement of \Lb\ branching fractions, the additional requirement 
$p_{T}(\Lb) > 6~\pgev$ was applied to allow easy comparison with other 
\Lb\ measurements at the Tevatron, which are only available above this 
threshold~\cite{Abulencia:2006df,Aaltonen:2008eu}. 
This additional requirement lowers the \Lb\ yields by about 20\%.
The raw fractions returned by the fit were corrected for the differences in selection efficiencies 
between different modes, which range from 8\% to 40\% for the 
measurements of $b$--mesons and \Lb\ branching 
fractions, respectively. These corrections were determined from detailed detector simulation
, with the following exceptions that were measured from data: 
the momentum-averaged relative isolation efficiency between \Bs\ 
and \Bd, $1.00 \pm 0.03$, has been determined from fully-reconstructed samples of
\Bs $\rightarrow J/\psi\phi$, 
and \Bd $\rightarrow J/\psi K^{*0}$ decays~\cite{Morello-Thesis};
the difference in efficiency for triggering on kaons and pions due to 
the different specific ionization in the COT (a $\approx 5$\% effect) was measured 
from a sample of $D^{+}\rightarrow K^{-}\pi^{+}\pi^{+}$ 
decays triggered on two tracks, using the unbiased third track~\cite{Acosta:2004ts}.
Possible differences in efficiency of the isolation requirement 
between \Bd\ and \Lb, and in the trigger efficiency between kaons and protons, 
were taken into account in the systematic uncertainties.
 
The dominant contributions to the systematic uncertainty are the
uncertainty on the combinatorial background model and
the uncertainty on the \dedx\ calibration and parameterization. 
Other contributions come from trigger efficiencies, physics background shape and kinematics,  
$b$--hadron masses and lifetimes, and the possible polarization of \Lb\ decays. 


The final results are listed in Table \ref{tab:BRs}. Absolute 
branching fractions are also quoted, by normalizing to world-average 
values of production fractions and 
\BR(\BdKpi)~\cite{PDG08,Aubert:2006fha}. 
The branching fraction measured for the \BsKpi\ mode is consistent 
with the previous upper 
limit ($< 5.6 \times 10^{-6}$ at 90\%~C.L.), based on a subsample of the current data 
\cite{Abulencia:2006psa}. 
This agrees with the prediction in Ref.~\cite{Williamson:2006hb}, 
but it is lower than most other predictions 
\cite{Beneke:2003zv,Sun:2002rn,Chiang:2008vc}.
The \Bspipi\ upper limit improves and supersedes the previous best 
limit~\cite{Abulencia:2006psa}.
The present measurement of $\BR(\BdKK)$ is in agreement with other existing measurements and has a 
similar resolution~\cite{PDG08},
but the resulting upper limit is weaker due to the observed central value.
The sensitivity to both \BdKK\ and \Bspipi\ is now 
close to the upper end of the theoretically expected 
range~\cite{Beneke:2003zv,Sun:2002rn,Li:2004ep,Ali:2007ff}.  
We also report the first branching fraction measurements of charmless 
$\Lambda_b$ decays. They are
significantly lower than the previous upper limit of $2.3 \times 
10^{-5}$~\cite{Acosta:2005ab}, and
in reasonable agreement with predictions~\cite{Mohanta:2000nk}, thus excluding the possibility of
large ($O(10^{2})$) enhancements from R-parity violating 
supersymmetric scenarios~\cite{Mohanta:2000za}. 
Their ratio can be determined directly from our data with greater 
accuracy than the individual values. For this purpose, the 
additional $p_{T} > 6~\pgev$ requirement is not necessary, and we can exploit 
the full sample size, obtaining
$\BR(\Lbppi)/\BR(\LbpK) =  0.66 \pm 0.14 \pm 0.08$, 
in good agreement with the predicted range 0.60--0.62~\cite{Mohanta:2000nk}.

In summary, we have searched for rare charmless decay modes of neutral
$b$--hadrons into pairs of charged hadrons in CDF data. 
We report the first observation of the modes \BsKpi, \Lbppi, and \LbpK,
and measure their relative branching fractions.
We set upper limits on the unobserved modes \BdKK\ and \Bspipi.
\begin{acknowledgments}
\input{ack_081222}

\end{acknowledgments}
 %
 %


\input{PRL_Bhh_BR_v30.bbl}

\end{document}

%% file: July_2008_Authors_Dec22_Rick.tex
\affiliation{Institute of Physics, Academia Sinica, Taipei, Taiwan 11529, Republic of China} 
\affiliation{Argonne National Laboratory, Argonne, Illinois 60439} 
\affiliation{University of Athens, 157 71 Athens, Greece} 
\affiliation{Institut de Fisica d'Altes Energies, Universitat Autonoma de Barcelona, E-08193, Bellaterra (Barcelona), Spain} 
\affiliation{Baylor University, Waco, Texas  76798} 
\affiliation{Istituto Nazionale di Fisica Nucleare Bologna, $^v$University of Bologna, I-40127 Bologna, Italy} 
\affiliation{Brandeis University, Waltham, Massachusetts 02254} 
\affiliation{University of California, Davis, Davis, California  95616} 
\affiliation{University of California, Los Angeles, Los Angeles, California  90024} 
\affiliation{University of California, San Diego, La Jolla, California  92093} 
\affiliation{University of California, Santa Barbara, Santa Barbara, California 93106} 
\affiliation{Instituto de Fisica de Cantabria, CSIC-University of Cantabria, 39005 Santander, Spain} 
\affiliation{Carnegie Mellon University, Pittsburgh, PA  15213} 
\affiliation{Enrico Fermi Institute, University of Chicago, Chicago, Illinois 60637}
\affiliation{Comenius University, 842 48 Bratislava, Slovakia; Institute of Experimental Physics, 040 01 Kosice, Slovakia} 
\affiliation{Joint Institute for Nuclear Research, RU-141980 Dubna, Russia} 
\affiliation{Duke University, Durham, North Carolina  27708} 
\affiliation{Fermi National Accelerator Laboratory, Batavia, Illinois 60510} 
\affiliation{University of Florida, Gainesville, Florida  32611} 
\affiliation{Laboratori Nazionali di Frascati, Istituto Nazionale di Fisica Nucleare, I-00044 Frascati, Italy} 
\affiliation{University of Geneva, CH-1211 Geneva 4, Switzerland} 
\affiliation{Glasgow University, Glasgow G12 8QQ, United Kingdom} 
\affiliation{Harvard University, Cambridge, Massachusetts 02138} 
\affiliation{Division of High Energy Physics, Department of Physics, University of Helsinki and Helsinki Institute of Physics, FIN-00014, Helsinki, Finland} 
\affiliation{University of Illinois, Urbana, Illinois 61801} 
\affiliation{The Johns Hopkins University, Baltimore, Maryland 21218} 
\affiliation{Institut f\"{u}r Experimentelle Kernphysik, Universit\"{a}t Karlsruhe, 76128 Karlsruhe, Germany} 
\affiliation{Center for High Energy Physics: Kyungpook National University, Daegu 702-701, Korea; Seoul National University, Seoul 151-742, Korea; Sungkyunkwan University, Suwon 440-746, Korea; Korea Institute of Science and Technology Information, Daejeon, 305-806, Korea; Chonnam National University, Gwangju, 500-757, Korea} 
\affiliation{Ernest Orlando Lawrence Berkeley National Laboratory, Berkeley, California 94720} 
\affiliation{University of Liverpool, Liverpool L69 7ZE, United Kingdom} 
\affiliation{University College London, London WC1E 6BT, United Kingdom} 
\affiliation{Centro de Investigaciones Energeticas Medioambientales y Tecnologicas, E-28040 Madrid, Spain} 
\affiliation{Massachusetts Institute of Technology, Cambridge, Massachusetts  02139} 
\affiliation{Institute of Particle Physics: McGill University, Montr\'{e}al, Qu\'{e}bec, Canada H3A~2T8; Simon Fraser University, Burnaby, British Columbia, Canada V5A~1S6; University of Toronto, Toronto, Ontario, Canada M5S~1A7; and TRIUMF, Vancouver, British Columbia, Canada V6T~2A3} 
\affiliation{University of Michigan, Ann Arbor, Michigan 48109} 
\affiliation{Michigan State University, East Lansing, Michigan  48824}
\affiliation{Institution for Theoretical and Experimental Physics, ITEP, Moscow 117259, Russia} 
\affiliation{University of New Mexico, Albuquerque, New Mexico 87131} 
\affiliation{Northwestern University, Evanston, Illinois  60208} 
\affiliation{The Ohio State University, Columbus, Ohio  43210} 
\affiliation{Okayama University, Okayama 700-8530, Japan} 
\affiliation{Osaka City University, Osaka 588, Japan} 
\affiliation{University of Oxford, Oxford OX1 3RH, United Kingdom} 
\affiliation{Istituto Nazionale di Fisica Nucleare, Sezione di Padova-Trento, $^w$University of Padova, I-35131 Padova, Italy} 
\affiliation{LPNHE, Universite Pierre et Marie Curie/IN2P3-CNRS, UMR7585, Paris, F-75252 France} 
\affiliation{University of Pennsylvania, Philadelphia, Pennsylvania 19104}
\affiliation{Istituto Nazionale di Fisica Nucleare Pisa, $^x$University of Pisa, $^y$University of Siena and $^z$Scuola Normale Superiore, I-56127 Pisa, Italy} 
\affiliation{University of Pittsburgh, Pittsburgh, Pennsylvania 15260} 
\affiliation{Purdue University, West Lafayette, Indiana 47907} 
\affiliation{University of Rochester, Rochester, New York 14627} 
\affiliation{The Rockefeller University, New York, New York 10021} 
\affiliation{Istituto Nazionale di Fisica Nucleare, Sezione di Roma 1, $^{aa}$Sapienza Universit\`{a} di Roma, I-00185 Roma, Italy} 

\affiliation{Rutgers University, Piscataway, New Jersey 08855} 
\affiliation{Texas A\&M University, College Station, Texas 77843} 
\affiliation{Istituto Nazionale di Fisica Nucleare Trieste/Udine, I-34100 Trieste, $^{bb}$University of Trieste/Udine, I-33100 Udine, Italy} 
\affiliation{University of Tsukuba, Tsukuba, Ibaraki 305, Japan} 
\affiliation{Tufts University, Medford, Massachusetts 02155} 
\affiliation{Waseda University, Tokyo 169, Japan} 
\affiliation{Wayne State University, Detroit, Michigan  48201} 
\affiliation{University of Wisconsin, Madison, Wisconsin 53706} 
\affiliation{Yale University, New Haven, Connecticut 06520} 
\author{T.~Aaltonen}
\affiliation{Division of High Energy Physics, Department of Physics, University of Helsinki and Helsinki Institute of Physics, FIN-00014, Helsinki, Finland}
\author{J.~Adelman}
\affiliation{Enrico Fermi Institute, University of Chicago, Chicago, Illinois 60637}
\author{T.~Akimoto}
\affiliation{University of Tsukuba, Tsukuba, Ibaraki 305, Japan}
\author{B.~\'{A}lvarez~Gonz\'{a}lez$^q$}
\affiliation{Instituto de Fisica de Cantabria, CSIC-University of Cantabria, 39005 Santander, Spain}
\author{S.~Amerio$^w$}
\affiliation{Istituto Nazionale di Fisica Nucleare, Sezione di Padova-Trento, $^w$University of Padova, I-35131 Padova, Italy} 

\author{D.~Amidei}
\affiliation{University of Michigan, Ann Arbor, Michigan 48109}
\author{A.~Anastassov}
\affiliation{Northwestern University, Evanston, Illinois  60208}
\author{A.~Annovi}
\affiliation{Laboratori Nazionali di Frascati, Istituto Nazionale di Fisica Nucleare, I-00044 Frascati, Italy}
\author{J.~Antos}
\affiliation{Comenius University, 842 48 Bratislava, Slovakia; Institute of Experimental Physics, 040 01 Kosice, Slovakia}
\author{G.~Apollinari}
\affiliation{Fermi National Accelerator Laboratory, Batavia, Illinois 60510}
% ...

\collaboration{CDF Collaboration\footnote{With visitors from $^a$University of Massachusetts Amherst, Amherst, Massachusetts 01003,
$^b$Universiteit Antwerpen, B-2610 Antwerp, Belgium, 
$^c$University of Bristol, Bristol BS8 1TL, United Kingdom,
$^d$Chinese Academy of Sciences, Beijing 100864, China, 
$^e$Istituto Nazionale di Fisica Nucleare, Sezione di Cagliari, 09042 Monserrato (Cagliari), Italy,
$^f$University of California Irvine, Irvine, CA  92697, 
$^g$University of California Santa Cruz, Santa Cruz, CA  95064, 
$^h$Cornell University, Ithaca, NY  14853, 
$^i$University of Cyprus, Nicosia CY-1678, Cyprus, 
$^j$University College Dublin, Dublin 4, Ireland,
$^k$University of Edinburgh, Edinburgh EH9 3JZ, United Kingdom, 
$^l$Universidad Iberoamericana, Mexico D.F., Mexico,
$^m$Queen Mary, University of London, London, E1 4NS, England,
$^n$University of Manchester, Manchester M13 9PL, England, 
$^o$Nagasaki Institute of Applied Science, Nagasaki, Japan, 
$^p$University of Notre Dame, Notre Dame, IN 46556,
$^q$University de Oviedo, E-33007 Oviedo, Spain, 
$^r$Texas Tech University, Lubbock, TX  79409, 
$^s$IFIC(CSIC-Universitat de Valencia), 46071 Valencia, Spain,
$^t$University of Virginia, Charlottesville, VA  22904,
$^u$Bergische Universit\"at Wuppertal, 42097 Wuppertal, Germany,
$^{cc}$On leave from J.~Stefan Institute, Ljubljana, Slovenia, 
}}
\noaffiliation

%% file: ack_081222.tex
We thank the Fermilab staff and the technical staffs of the participating 
institutions for their vital contributions. This work was supported by 
the U.S. Department of Energy and National Science Foundation; the 
Italian Istituto Nazionale di Fisica Nucleare; the Ministry of Education, 
Culture, Sports, Science and Technology of Japan; the Natural Sciences 
and Engineering Research Council of Canada; the National Science Council 
of the Republic of China; the Swiss National Science Foundation; the A.P. 
Sloan Foundation; the Bundesministerium f\"ur Bildung und Forschung, 
Germany; the Korean Science and Engineering Foundation and the Korean 
Research Foundation; the Science and Technology Facilities Council and the Royal 
Society, UK; the Institut National de Physique Nucleaire et 
Physique des Particules/CNRS; the Russian Foundation for Basic Research; the
Ministerio de Ciencia e Innovaci\'{o}n, and Programa Consolider-Ingenio 2010, Spain;
the Slovak R\&D Agency; and the Academy of Finland.

%% file: PRL_Bhh_BR_V30.bbl
\begin{thebibliography}{37}
\expandafter\ifx\csname natexlab\endcsname\relax\def\natexlab#1{#1}\fi
\expandafter\ifx\csname bibnamefont\endcsname\relax
 \def\bibnamefont#1{#1}\fi
\expandafter\ifx\csname bibfnamefont\endcsname\relax
 \def\bibfnamefont#1{#1}\fi
\expandafter\ifx\csname citenamefont\endcsname\relax
 \def\citenamefont#1{#1}\fi
\expandafter\ifx\csname url\endcsname\relax
 \def\url#1{\texttt{#1}}\fi
\expandafter\ifx\csname urlprefix\endcsname\relax\def\urlprefix{URL }\fi
\providecommand{\bibinfo}[2]{#2}
\providecommand{\eprint}[2][]{\url{#2}}

\bibitem[{\citenamefont{{B. Aubert \em et al.}}(2007)}]{Aubert:2006fha}
\bibinfo{author}{\bibnamefont{{B. Aubert \em et al.}}}
(\bibinfo{collaboration}{{\mbox{\sl B\hspace{-0.4em} {\small\sl
A}\hspace{-0.37em} \sl B\hspace{-0.4em} {\small\sl A\hspace{-0.02em}R}}}
Collaboration}), \bibinfo{journal}{Phys. Rev. D}
\textbf{\bibinfo{volume}{75}}, \bibinfo{pages}{012008}
(\bibinfo{year}{2007});
(\bibinfo{collaboration}{Belle Collaboration}), \bibinfo{journal}{Phys. Rev.
Lett.} \textbf{\bibinfo{volume}{99}}, \bibinfo{pages}{121601}
(\bibinfo{year}{2007});
\bibinfo{author}{\bibnamefont{{A. Bornheim \em et al.}}}
(\bibinfo{collaboration}{CLEO Collaboration}), \bibinfo{journal}{Phys. Rev.
D} \textbf{\bibinfo{volume}{68}}, \bibinfo{pages}{052002}
(\bibinfo{year}{2003}).

\bibitem[{\citenamefont{Fleischer}(1999)}]{Fleischer:1999pa}
\bibinfo{author}{\bibfnamefont{R.}~\bibnamefont{Fleischer}},
 \bibinfo{journal}{Phys. Lett. B} \textbf{\bibinfo{volume}{459}},
 \bibinfo{pages}{306} (\bibinfo{year}{1999});
\bibinfo{author}{\bibfnamefont{A.}~\bibnamefont{Soni}} \bibnamefont{and}
 \bibinfo{author}{\bibfnamefont{D.~A.} \bibnamefont{Suprun}},
 \bibinfo{journal}{Phys. Rev. D} \textbf{\bibinfo{volume}{75}},
 \bibinfo{pages}{054006} (\bibinfo{year}{2007}).

\bibitem[{\citenamefont{{A. Abulencia \em et al.}}(2006)}]{Abulencia:2006psa}
\bibinfo{author}{\bibnamefont{{A. Abulencia \em et al.}}}
 (\bibinfo{collaboration}{CDF Collaboration}), \bibinfo{journal}{Phys. Rev.
 Lett.} \textbf{\bibinfo{volume}{97}}, \bibinfo{pages}{211802}
 (\bibinfo{year}{2006}).

\bibitem[{\citenamefont{Gronau and Rosner}(2000)}]{Gronau:2000md}
\bibinfo{author}{\bibfnamefont{M.}~\bibnamefont{Gronau}} \bibnamefont{and}
 \bibinfo{author}{\bibfnamefont{J.~L.} \bibnamefont{Rosner}},
 \bibinfo{journal}{Phys. Lett. B} \textbf{\bibinfo{volume}{482}},
 \bibinfo{pages}{71} (\bibinfo{year}{2000}).

\bibitem[{\citenamefont{Sun et~al.}(2003)\citenamefont{Sun, Zhu, and
 Du}}]{Sun:2002rn}
\bibinfo{author}{\bibfnamefont{J.-F.} \bibnamefont{Sun}},
 \bibinfo{author}{\bibfnamefont{G.-H.} \bibnamefont{Zhu}}, \bibnamefont{and}
 \bibinfo{author}{\bibfnamefont{D.-S.} \bibnamefont{Du}},
 \bibinfo{journal}{Phys. Rev. D} \textbf{\bibinfo{volume}{68}},
 \bibinfo{pages}{054003} (\bibinfo{year}{2003}).

\bibitem[{\citenamefont{Beneke and Neubert}(2003)}]{Beneke:2003zv}
\bibinfo{author}{\bibfnamefont{M.}~\bibnamefont{Beneke}} \bibnamefont{and}
 \bibinfo{author}{\bibfnamefont{M.}~\bibnamefont{Neubert}},
 \bibinfo{journal}{Nucl. Phys.} \textbf{\bibinfo{volume}{B675}},
 \bibinfo{pages}{333} (\bibinfo{year}{2003}).

\bibitem[{\citenamefont{{A. Ali \em et al.}}(2007)}]{Ali:2007ff}
\bibinfo{author}{\bibnamefont{{A. Ali \em et al.}}}, \bibinfo{journal}{Phys.
 Rev. D} \textbf{\bibinfo{volume}{76}}, \bibinfo{pages}{074018}
 (\bibinfo{year}{2007});
\bibinfo{author}{\bibfnamefont{X.-Q.} \bibnamefont{Yu}},
 \bibinfo{author}{\bibfnamefont{Y.}~\bibnamefont{Li}}, \bibnamefont{and}
 \bibinfo{author}{\bibfnamefont{C.-D.} \bibnamefont{Lu}},
 \bibinfo{journal}{Phys. Rev. D} \textbf{\bibinfo{volume}{71}},
 \bibinfo{pages}{074026} (\bibinfo{year}{2005}); {{\em Erratum ibid.}} 
\textbf{\bibinfo{volume}{72}}, \bibinfo{pages}{119903 (E)} (\bibinfo{year}{2005})

\bibitem[{\citenamefont{Buras et~al.}(2004)\citenamefont{Buras, Fleischer,
 Recksiegel, and Schwab}}]{Buras:2004ub}
\bibinfo{author}{\bibfnamefont{A.~J.} \bibnamefont{Buras}},
 \bibinfo{author}{\bibfnamefont{R.}~\bibnamefont{Fleischer}},
 \bibinfo{author}{\bibfnamefont{S.}~\bibnamefont{Recksiegel}},
 \bibnamefont{and} \bibinfo{author}{\bibfnamefont{F.}~\bibnamefont{Schwab}},
 \bibinfo{journal}{Nucl. Phys.} \textbf{\bibinfo{volume}{B697}},
 \bibinfo{pages}{133} (\bibinfo{year}{2004}).

\bibitem[{\citenamefont{Mohanta et~al.}(2001)\citenamefont{Mohanta, Giri, and
 Khanna}}]{Mohanta:2000nk}
\bibinfo{author}{\bibfnamefont{R.}~\bibnamefont{Mohanta}},
 \bibinfo{author}{\bibfnamefont{A.~K.} \bibnamefont{Giri}}, \bibnamefont{and}
 \bibinfo{author}{\bibfnamefont{M.~P.} \bibnamefont{Khanna}},
 \bibinfo{journal}{Phys. Rev. D} \textbf{\bibinfo{volume}{63}},
 \bibinfo{pages}{074001} (\bibinfo{year}{2001}).

\bibitem[{C-c()}]{C-conjugate}
\emph{\bibinfo{title}{{\rm Throughout this paper, C-conjugate modes are implied
 and branching fractions indicate CP-averages.}}}

\bibitem[{\citenamefont{{D. Acosta \em et
 al.}}(2005{\natexlab{a}})}]{Acosta:2004yw}
\bibinfo{author}{\bibnamefont{{D. Acosta \em et al.}}}
 (\bibinfo{collaboration}{CDF Collaboration}), \bibinfo{journal}{Phys. Rev. D}
 \textbf{\bibinfo{volume}{71}}, \bibinfo{pages}{032001}
 (\bibinfo{year}{2005}{\natexlab{a}});
%
\bibinfo{author}{\bibfnamefont{A.}~\bibnamefont{Sill}}
 (\bibinfo{collaboration}{CDF Collaboration}), \bibinfo{journal}{Nucl.
 Instrum. Methods A} \textbf{\bibinfo{volume}{447}}, \bibinfo{pages}{1}
 (\bibinfo{year}{2000});
\bibinfo{author}{\bibnamefont{{A. Affolder \em et al.}}},
 \bibinfo{journal}{Nucl. Instrum. Methods A} \textbf{\bibinfo{volume}{453}},
 \bibinfo{pages}{84} (\bibinfo{year}{2000});
%
\bibinfo{author}{\bibnamefont{{T. Affolder \em et al.}}},
 \bibinfo{journal}{Nucl. Instrum. Methods A} \textbf{\bibinfo{volume}{526}},
 \bibinfo{pages}{249} (\bibinfo{year}{2004}).

\bibitem[{CDF()}]{CDF-coordinates}
\emph{\bibinfo{title}{{\em CDF II uses a cylindrical coordinate system in which
 $\phi$ is the azimuthal angle, $r$ is the radius from the nominal beam line,
 and $z$ points in the proton beam direction, with the origin at the center of
 the detector. The transverse plane is the plane perpendicular to the $z$
 axis.}}}

\bibitem[{\citenamefont{{E. Thomson \em et al.}}(2002)}]{Thomson:2002xp}
\bibinfo{author}{\bibnamefont{{E. Thomson \em et al.}}}, \bibinfo{journal}{IEEE
 Trans. Nucl. Sci.} \textbf{\bibinfo{volume}{49}}, \bibinfo{pages}{1063}
 (\bibinfo{year}{2002});
\bibinfo{author}{\bibnamefont{{R.~Downing \em et al.}}},
 \bibinfo{journal}{Nucl. Instrum. Methods A} \textbf{\bibinfo{volume}{570}},
 \bibinfo{pages}{36} (\bibinfo{year}{2007}).

\bibitem[{\citenamefont{{B. Ashmanskas \em et al.}}(2004)}]{Ashmanskas:2003gf}
\bibinfo{author}{\bibnamefont{{B. Ashmanskas \em et al.}}},
 \bibinfo{journal}{Nucl. Instrum. Methods A} \textbf{\bibinfo{volume}{518}},
 \bibinfo{pages}{532} (\bibinfo{year}{2004}).

\bibitem[{Iso()}]{Isolation}
\emph{\bibinfo{title}{{\em Isolation is defined as
 $I_{B}=p_T(B)/(p_T(B)+\sum_ip_{Ti})$, where $p_T(B)$ is the transverse
 momentum of the $B$ candidate, and the sum runs over all other tracks within
 a cone of radius 1, in $\eta$-$\phi$ space around the $B$ flight-direction.}}}

\bibitem[{\citenamefont{Punzi}(2003)}]{Punzi:2003bu}
\bibinfo{author}{\bibfnamefont{G.}~\bibnamefont{Punzi}},
 \bibinfo{journal}{eConf} \textbf{\bibinfo{volume}{C030908}},
 \bibinfo{pages}{MODT002} (\bibinfo{year}{2003}),
 \eprint{ArXiv:physics/0308063}.

\bibitem[{\citenamefont{{M.J. Morello}}(2007)}]{Morello-Thesis}
\bibinfo{author}{\bibnamefont{{M.J. Morello}}}, \bibinfo{journal}{Ph.D. Thesis,
 Scuola Normale Superiore, Pisa, Fermilab Report No. FERMILAB-THESIS-2007-57}
 (\bibinfo{year}{2007}).

\bibitem[{\citenamefont{{H. Albrecht \em et al.}}(1990)}]{Albrecht:1990am}
\bibinfo{author}{\bibnamefont{{H. Albrecht \em et al.}}}
 (\bibinfo{collaboration}{ARGUS Collaboration}), \bibinfo{journal}{Phys. Lett.
 B} \textbf{\bibinfo{volume}{241}}, \bibinfo{pages}{278}
 (\bibinfo{year}{1990}).

\bibitem[{\citenamefont{Baracchini and Isidori}(2006)}]{Baracchini:2005wp}
\bibinfo{author}{\bibfnamefont{E.}~\bibnamefont{Baracchini}} \bibnamefont{and}
 \bibinfo{author}{\bibfnamefont{G.}~\bibnamefont{Isidori}},
 \bibinfo{journal}{Phys. Lett. B} \textbf{\bibinfo{volume}{633}},
 \bibinfo{pages}{309} (\bibinfo{year}{2006}).

\bibitem[{\citenamefont{{T. Aaltonen \em et al.}}(2008)}]{Aaltonen:2008eu}
\bibinfo{author}{\bibnamefont{{T. Aaltonen \em et al.}}}
 (\bibinfo{collaboration}{CDF Collaboration}), \bibinfo{journal}{submitted to
 Phys. Rev. D}  (\bibinfo{year}{2008}), \eprint{{arXiv:0810.3213}}.

\bibitem[{\citenamefont{{C. Amsler \em et al.}}(2008)}]{PDG08}
\bibinfo{author}{\bibnamefont{{C. Amsler \em et al.}}}, \bibinfo{journal}{Phys.
 Lett. B} \textbf{\bibinfo{volume}{667}}, \bibinfo{pages}{1}
 (\bibinfo{year}{2008}).

\bibitem[{\citenamefont{Feldman and Cousins}(1998)}]{Feldman:1997qc}
\bibinfo{author}{\bibfnamefont{G.~J.} \bibnamefont{Feldman}} \bibnamefont{and}
 \bibinfo{author}{\bibfnamefont{R.~D.} \bibnamefont{Cousins}},
 \bibinfo{journal}{Phys. Rev. D} \textbf{\bibinfo{volume}{57}},
 \bibinfo{pages}{3873} (\bibinfo{year}{1998}).

\bibitem[{\citenamefont{{A. Abulencia \em et al.}}(2007)}]{Abulencia:2006df}
\bibinfo{author}{\bibnamefont{{A. Abulencia \em et al.}}}
 (\bibinfo{collaboration}{CDF Collaboration}), \bibinfo{journal}{Phys. Rev.
 Lett.} \textbf{\bibinfo{volume}{98}}, \bibinfo{pages}{122002}
 (\bibinfo{year}{2007}).

\bibitem[{\citenamefont{{D. Acosta \em et
 al.}}(2005{\natexlab{b}})}]{Acosta:2004ts}
\bibinfo{author}{\bibnamefont{{D. Acosta \em et al.}}}
 (\bibinfo{collaboration}{CDF Collaboration}), \bibinfo{journal}{Phys. Rev.
 Lett.} \textbf{\bibinfo{volume}{94}}, \bibinfo{pages}{122001}
 (\bibinfo{year}{2005}{\natexlab{b}}).

\bibitem[{\citenamefont{Williamson and Zupan}(2006)}]{Williamson:2006hb}
\bibinfo{author}{\bibfnamefont{A.~R.} \bibnamefont{Williamson}}
 \bibnamefont{and} \bibinfo{author}{\bibfnamefont{J.}~\bibnamefont{Zupan}},
 \bibinfo{journal}{Phys. Rev. D} \textbf{\bibinfo{volume}{74}},
 \bibinfo{pages}{014003} (\bibinfo{year}{2006}); {{\em Erratum ibid.}} 
\textbf{\bibinfo{volume}{72}}, \bibinfo{pages}{039991 (E)} (\bibinfo{year}{2006})

\bibitem[{\citenamefont{Chiang et~al.}(2008)\citenamefont{Chiang, Gronau, and
 Rosner}}]{Chiang:2008vc}
\bibinfo{author}{\bibfnamefont{C.-W.} \bibnamefont{Chiang}},
 \bibinfo{author}{\bibfnamefont{M.}~\bibnamefont{Gronau}}, \bibnamefont{and}
 \bibinfo{author}{\bibfnamefont{J.~L.} \bibnamefont{Rosner}},
 \bibinfo{journal}{Phys. Lett. B} \textbf{\bibinfo{volume}{664}},
 \bibinfo{pages}{169} (\bibinfo{year}{2008}).

\bibitem[{\citenamefont{Li et~al.}(2004)\citenamefont{Li, Lu, Xiao, and
 Yu}}]{Li:2004ep}
\bibinfo{author}{\bibfnamefont{Y.}~\bibnamefont{Li}},
 \bibinfo{author}{\bibfnamefont{C.-D.} \bibnamefont{Lu}},
 \bibinfo{author}{\bibfnamefont{Z.-J.} \bibnamefont{Xiao}}, \bibnamefont{and}
 \bibinfo{author}{\bibfnamefont{X.-Q.} \bibnamefont{Yu}},
 \bibinfo{journal}{Phys. Rev. D} \textbf{\bibinfo{volume}{70}},
 \bibinfo{pages}{034009} (\bibinfo{year}{2004}).

\bibitem[{\citenamefont{{D. Acosta \em et
 al.}}(2005{\natexlab{c}})}]{Acosta:2005ab}
\bibinfo{author}{\bibnamefont{{D. Acosta \em et al.}}}
 (\bibinfo{collaboration}{CDF Collaboration}), \bibinfo{journal}{Phys. Rev. D}
 \textbf{\bibinfo{volume}{72}}, \bibinfo{pages}{051104}
 (\bibinfo{year}{2005}{\natexlab{c}}).

\bibitem[{\citenamefont{Mohanta}(2001)}]{Mohanta:2000za}
\bibinfo{author}{\bibfnamefont{R.}~\bibnamefont{Mohanta}},
 \bibinfo{journal}{Phys. Rev. D} \textbf{\bibinfo{volume}{63}},
 \bibinfo{pages}{056006} (\bibinfo{year}{2001}).

\end{thebibliography}
